\documentclass[prl,showpacs,twocolumn,aps]{revtex4}
\begin{document}
\draft \title{A variational coupled-cluster study of magnon-density-wave
excitations in quantum antiferromagnets} 
\author{Y. Xian}
\address{School of Physics and Astronomy, The University of
Manchester, Manchester M13 9PL, UK} \date{\today } 

\begin{abstract}

We extend recently proposed variational coupled-cluster method to
describe excitation states of quantum antiferromagnetic bipartite lattices.
We reproduce the spin-wave excitations (i.e., magnons with spin $\pm 1$).
In addition, we obtain a new, spin-zero excitation (magnon-density waves)
which has been missing in all existing spin-wave theories. Within our
approximation, this magnon-density-wave excitation has a nonzero energy gap
in a cubic lattice and is gapless in a square lattice, similar to those
of charge-density-wave excitations (plasmons) in quantum electron gases.

\end{abstract}


\maketitle

From many-body physics view, energy spectra of low-lying
excitation states of a quantum many-body interacting
system are mainly determined by the correlations included
in its ground state. Feynman's description of the phonon-roton
excitations in a Helium-4 quantum liquid is a well-known 
example \cite{fey,fee}. For the ground state properties,
the method of correlated basis functions (CBF) has 
produced satisfactory results. Combining with Feynman's excitation theory,
we therefore have an analytical, systematic method capable of describing most
states of the Helium-4 quantum liquid at low temperature \cite{fee}.

The coupled-cluster method (CCM) \cite{hhc,ck,ciz} is another
successful microscopic many-body theory which has produced
many most competitive results for the ground state properties of 
many atoms, molecules, electron gas, and quantum spin systems \cite{bb}.
In the traditional CCM, however, the bra and ket states 
are not hermitian to one another \cite{arp}.
We recently extended the CCM to a variational formalism \cite{yx1,yx2},
in which bra and ket states are hermitian to one another. In our
analysis, a close relation with the CBF method was found and exploited by
using diagrammatic techniques. In application to quantum antiferromagnets, 
our calculations in a simple approximation reproduced the ground state
properties of the spin-wave theory (SWT) \cite{and}. Improvements over
SWT by higher-order calculations were also obtained \cite{yx2}.
Here we extend this variational method to describe excitation 
states. Inspired by the close relation to the CBF method, we are 
able to obtain the energy spectra of quasiparticle-density-wave
excitations by following Feynman, as well as the usual quasiparticle 
excitation spectra as in the traditional
CCM. For the antiferromagnetic models, these quasiparticle excitations
correspond to the well known spin-wave excitations (magnons) with spin $\pm1$;
the new, spin-zero excitations are identified as magnon-density-wave
excitations whose energy spectra share some characteristics with 
that of charge-density-wave 
excitations (plasmons) in quantum electron gases \cite{dp}. We like to 
emphasize that these  spin-zero excitations have been missing 
in all spin-wave theories, including the more recently modified 
spin-wave theories \cite{aa,ht,mt}. Magnon-density fluctuations
in Heisenberg  ferromagnets at low temperature were investigated in the
sixties by calculating the longitudinal susceptibility \cite{yvg,vlp}.
Calculations for the longitudinal spin fluctuations due to multi-spin-wave
excitations in antiferromagnets were also performed by a bosonization
scheme \cite{bm} and by finite-size (classical) dynamic 
simulations \cite{bl}. Both our analysis and our results are different
to these calculations and more discussion will be given in conclusion.

The Heisenberg Hamiltonian is given by
\begin{equation}
H=\frac 12\sum_{l,\rho }H_{l,l+n}=\frac 14\sum_{l,n}
\left(2s_l^zs_{l+n}^z+s_l^{+}s_{l+n}^{-}+s_l^{-}s_{l+n}^{+}\right),
\end{equation}
where the index $l$ runs over all $N$
bipartite lattice sites, $n$ runs over all $z$ nearest-neighbor
sites, $s^z$ and $s^\pm$ are spin operators. 
As shown in Refs.~ 8 and 9, we use Coester representation for
the ground-state $|\Psi_g\rangle$ of Eq.~(1),
\begin{equation}
|\Psi_g\rangle=e^S|\Phi\rangle,\quad S=\sum_I F_I C^\dag_I
\end{equation}
with its hermitian conjugate $\langle\tilde\Psi_g|=\langle\Phi|e^{\tilde S}$,
$\tilde S=\sum_I \tilde F_I C_I$ as the bra ground-state. In Eq.~(2),
the model state $|\Phi\rangle$ is given by the N\'eel state 
with alternating spin up and down sublattices, and $C_I^\dagger$ with
nominal index $I$ is given by the spin-flip operators,
\begin{equation}
\sum_I F_I C^\dag_I= \sum_{k=1}^{N/2}
\sum_{i_1...,j_1...}f_{i_1...,j_1...}
\frac{s_{i_1}^{-}...s_{i_k}^{-}s_{j_1}^{+}...s_{j_k}^{+}}{(2s)^k},
\end{equation}
with $s$ as spin quantum number. The bra state
operators are given by the corresponding hermitian conjugate of
Eq.~(3), using notation $\tilde F_I = \tilde f_{i_1...,j_1...}$ for
the bra-state coefficients. As before, we have used index $i$ exclusively
for the spin-up sublattice of the N\'eel state and index $j$ for
corresponding the spin-down sublattice. The coefficients $\{F_I,\tilde
F_I\}$ are then determined by the usual variational equations as
\begin{equation}
\frac{\delta\langle H\rangle}{\delta \tilde F_I} =
\frac{\delta\langle H\rangle}{\delta F_I} = 0,
\quad
  \langle H\rangle \equiv 
  \frac{\langle\tilde\Psi_g| H|\Psi_g\rangle}
       {\langle\tilde\Psi_g|\Psi_g\rangle}.
\end{equation}
The important bare distribution functions, $g_I\equiv\langle C_I\rangle$
and $\tilde g_I\equiv\langle C^\dagger_I\rangle$, can be expressed in
a self-consistency equation as 
\begin{equation}
g_I = G(\tilde g_J,F_J),\quad \tilde g_I = G(g_J, \tilde F_J),
\end{equation}
where $G$ is a function containing up to linear terms in $\tilde g_J$
(or $g_J$) and finite order terms in $F_J$ (or $\tilde F_J$).
In Refs.~8 and 9, as a demonstration, we considered a truncation
approximation in which the correlation operators $S$ and
$\tilde S$ of Eqs.~(2-3) retain only the two-spin-flip
operators as
\begin{equation}
S\approx \sum_{ij}f_{ij}\frac{s_i^{-}s_j^{+}}{2s},\quad 
\tilde S\approx \sum_{ij}\tilde f_{ij}\frac{s_i^{+}s_j^{-}}{2s}. 
\end{equation}
The spontaneous magnetization (order-parameter) in this two-spin-flip
approximation is obtained by  calculating the one-body
density function as
\begin{equation}
\langle s^z_i\rangle =s-\rho_i,\quad \rho_i = \sum_j\rho_{ij}
 = \sum_j f_{ij}\tilde g_{ij},
\end{equation}
where $\rho_i = \rho$ due to translational invariance of the
lattice system. As demonstrated in Ref.~9, contributions to one-body
density function of Eq.~(7) can be represented by diagrams. The
results of SWT are
reproduced by resuming all diagrams without any so-called Pauli lines
representing Pauli exclusion principle manifested by $(s^\pm)^2=0$ for
$s=1/2$. For example, the one-body bare distribution function 
$\tilde g_{ij}=\langle s^-_is^+_j\rangle/(2s)$ in this approximation 
is given by, after the sublattice Fourier transformation,
\begin{equation}
\tilde g_q \approx \frac{\tilde f_q}{1-\tilde f_qf_q},
\end{equation}
where $\tilde f_q$ is the Fourier component of correlation coefficient
$\tilde f_{ij}$ with $q$ restricted in the magnetic zone etc. Variational
Eq.~(4) reproduces the SWT result for this coefficient as
\begin{equation}
f_q=\tilde f_q = \frac{1}{\gamma_q}\left[\sqrt{1-(\gamma_q)^2}
-1\right],\quad
\gamma_q = \frac1z\sum_n e^{i\bf{q\cdot r_n}}.
\end{equation}
Finally, the two-body distribution functions 
$\tilde g_{ij,i'j'}=\langle s^-_is^+_js^-_{i'}s^+_{j'}\rangle/(2s)^2$
is approximated by
\begin{equation}
\tilde g_{ij,i'j'}\approx \tilde g_{ij}\tilde g_{i'j'}
 + \tilde g_{ij'}\tilde g_{i'j}\;.
\end{equation}
To go beyond SWT, we need to consider Pauli principle as
mentioned earlier by including diagrams with the so-called
Pauli lines. Details were described in Ref.~9. In the followings,
we will use approximations Eq.~(6-10) to discuss excitation states.

Order-parameter of Eq.~(7) can also be calculated through two-body
functions as
\begin{equation}
(\langle s^z_i\rangle)^2=\frac{\langle\tilde\Psi_g|(s^z_a)^2|\Psi_g\rangle}
{\langle\tilde\Psi_g|\Psi_g\rangle},
\end{equation}
where $s^z_a= \sum_l(-1)^l s^z_l/N$ is the staggered spin operator \cite{ht}.
Eq.~(11) is in fact the sum rule for the two-body distribution function
as in CBF \cite{fee}. This can seen by introducing (staggered) 
magnon-density operator $\hat n_i$ as
\begin{equation}
2\hat n_i = 2s-s^z_i+\frac1z\sum_{n=1}^z s^z_{i+n},
\end{equation}
where as before summation over $n$ is over all $z$ nearest neighbors.
Hence, the sum rule for the one-body function is simply
$\frac2N\sum_i\langle \hat n_i\rangle=\rho$. The two-body
Eq.~(11) can now be written in the following familiar sum rule equation as
\begin{equation}
\frac2N\sum_{i'=1}^{N/2}\langle \hat n_i\hat n_{i'}\rangle=\rho\rho_i
 = \rho^2,
\end{equation} 
where in the last equation, translational invariant property $\rho_i=\rho$
has been used. In the approximation of Eq.~(6-10), we find that this sum
rule is obeyed in both cubic and square lattices in the limit
$N\rightarrow\infty$. In particular, we find that
$(\frac2N\sum_{i'}\langle \hat n_i\hat n_{i'}\rangle - \rho^2)
\propto 1/N$ in a cubic lattice and $\propto (\ln N)/N$ in a square 
lattice \cite{yx3}. These asymptotic properties are important in the
corresponding excitation states as 
discussed later. However, Eq.~(13) is violated in the one-dimensional
model, showing the deficiency of the two-spin-flip approximation of Eq.~(6)
for the one-dimensional model. We therefore leave more detailed discussion
and a possible cure elsewhere \cite{yx3} and focus on the cubic and square
lattices in the followings, using approximations of Eqs.~(6-10).

For the quasiparticle excitations, briefly, we follow
Emrich in the traditional CCM \cite{emr,bpx} and
write excitation ket-state $|\Psi_e\rangle$ involving only 
$C^\dagger_I$ operators as
\begin{equation}
|\Psi_e\rangle = X|\Psi_g\rangle = Xe^S |\Phi\rangle,\quad
X=\sum_I x_IC^\dagger_I,
\end{equation}
but, unlike the traditional CCM, our bra excitation state is the
corresponding hermitian conjugate $\langle\tilde\Psi_e| = 
\langle\tilde \Psi_g|\tilde X = \langle\Phi|e^{\tilde S}\tilde X$.
Choosing a single spin-flip operator $C^\dagger_I=s^-_i$,
we have $X\approx \sum_i x_i s^-_i$ with coefficient chosen as
$x_i =x_i(q)=\sqrt{\frac2N}e^{i\bf{q\cdot r_i}}$ to define a linear
momentum $\bf q$. The state of Eq.~(14) has therefore 
spin $s^z_{\rm total}=-1$. The energy difference between the 
above excitation state and the variational ground state of Eqs.~(2-4),
$\epsilon_q = \langle\tilde\Psi_g|\tilde XHX|\Psi_g\rangle /
\langle\tilde\Psi_e|\Psi_e\rangle -\langle H\rangle$, 
can be derived as \cite{yx3}, to the order of $(2s)$,
\begin{equation}
 \epsilon_q\approx sz\frac{1+\rho_q+\gamma_q \tilde g_q}{1+\rho_q}=
   sz\sqrt{1-(\gamma_q)^2},
\end{equation}
where $\rho_q=f_q\tilde g_q$ and we have used Eqs.~(6-10). 
This agrees with SWT \cite{and} in this order. Spectrum of Eq.~(15)
is gapless in all dimensions because $\epsilon_q \propto q$ as 
$q\rightarrow0$. Similar calculations using spin-flip operators 
$C^\dagger_I=s^\dagger_j$ of the $j$-sublattice in Eq.~(14) will produce
the same spectrum as Eq.~(15) except that the corresponding excitation
state has spin $s^z_{\rm total}=+1$. These excitations are often
referred to as magnons \cite{and}.

We next consider the magnon-density-wave excitations. Quasiparticles
such as magnons in general interact with one another, thus producing
quasiparticle density fluctuations. The interaction potential
for magnons is described by the term $s^z_ls^z_{l+n}$ in the Hamiltonian
of Eq.~(1). Following Feynman as in CBF \cite{fey,fee}, we write our 
general quasiparticle-density-wave
excitation state as, using magnon-density operator as defined in Eq.~(12),
\begin{equation}
|\Psi_e^0\rangle = X^0_q|\Psi_g\rangle, \quad
X^0_q = \sum_i x_i(q) \hat n_i, \quad q>0
\end{equation}
and its hermitian counterparts for the bra state,
$\langle\tilde \Psi_e^0| = \langle\tilde\Psi_g|\tilde X^0_q$. The
coefficient $x_i(q)=\sqrt{\frac2N}e^{i\bf{q\cdot r_i}}$ as before.
The condition $q>0$ in Eq.~(16) ensures the orthogonality between this
excited state with the ground state. We notice that the density operator 
$\hat n_i$ in Eq.~(16) is a hermitian operator. This property can be
used to derive a double commutation formula for the energy 
difference between the above excitation state and the variational
ground states of Eqs.~(2-4) as,
\begin{equation}
\epsilon^0_q =\frac{\langle\tilde\Psi_g|\tilde X^0HX^0|\Psi_g\rangle}
{\langle\tilde\Psi_e^0|\Psi_e^0\rangle}-\langle H\rangle= 
\frac{N(q)}{S^0(q)},\quad q>0
\end{equation}
where $N(q)\equiv\langle[\tilde X^0_q,[H,X^0_q]]\rangle/2$,
$S^0(q)\equiv\langle\tilde X^0_qX^0_q\rangle$ is the structure
function, and notation $\langle \cdots\rangle$ is the ground state
expectation value as before. Details will be given elsewhere \cite{yx3}.
The double commutator $N(q)$ and structure function $S^0(q)$ can be
straightforwardly calculated as, using approximations of Eqs.~(6-10),
\begin{equation}
N(q) =-\frac{sz}{2}\sum_{q'}
(\gamma_{q'}+\gamma_q\gamma_{q-q'})\tilde g_{q'},
\end{equation}
and
\begin{equation}
S^0(q) = \frac14(1+\gamma^2_q)\rho + \frac14 \sum_{q'}[(1+
 \gamma^2_q)\rho_{q'}\rho_{q-q'}+
 2\gamma_q \tilde g_{q'}\tilde g_{q-q'}],
\end{equation}
where $\rho_q\equiv f_q\tilde g_q$ as before, $f_q$ and $\tilde g_q$
are as given by Eqs.~(8) and (9). Substituting Eqs.~(18) and (19) into Eq.~(17),
we can then calculate energy spectrum $\epsilon^0_q$ numerically.
We notice that Eq.~(19) is closely related to the the sum rule Eq.~(13)
which corresponds to $S^0(q)$ at $q=0$. It is not difficult to see from Eq.~(18), 
$N(q)$ has a nonzero, finite value for all values of $q$. Any special feature
such as gapless in the spectrum $\epsilon^0_q$ therefore comes from the
the structure function of Eq.~(19), and hence is determined by 
the asymptotic behaviors of the sum rule Eq.~(13) mentioned earlier.
For a cubic lattice, we find that the spectrum $\epsilon^0_q$ has
a nonzero gap everywhere. The minimum gap is $\epsilon^0_q\approx 0.99sz$ at
$q\rightarrow0$. This gap is about the same as the largest magnon 
energy, $\epsilon_q = sz$ at ${\bf q}=(\pi/2,\pi/2,\pi/2)$
from Eq.~(15). At ${\bf q}=(\pi/2,\pi/2,\pi/2)$,
we have the largest energy $\epsilon^0_q \approx 2.92sz$. This
is nearly three magnons' energy at this value of $\bf q$. At ${\bf q}=(\pi,0,0)$,
we obtain $\epsilon^0_q\approx2.56sz$. 

For the square lattice the structure function $S^0(q)$ of Eq.~(19)
has a logarithmic behavior $\ln q$ as $q\rightarrow0$. This is not surprising as
discussed earlier in the sum rule Eq.~(13), where occurs the asymptotic behavior
$(\ln N)/N$ as $N\rightarrow\infty$. For small values of $q$, $N(q)$ approaches
to a finite value, $N(q)\approx 0.275sz$ as $q\rightarrow0$. The corresponding energy
spectrum of Eq.~(17) is therefore gapless as $q\rightarrow 0$. We plot the spectrum
for the values of ${\bf q}$ between $(0.01\pi,\;0)$ to $(\pi,\;0)$ in Fig.~1, 
together with the corresponding magnon energies for comparison. As can be seen
from Fig.~1, magnon-density-wave energy is always larger than the corresponding
magnon energy. At small values of $q$ ($q<0.05\pi$), we find a good 
approximation by numerical calculations for the structure function,
$S^0(q)\approx 0.24-0.16\ln q$. The energy spectrum of Eq.~(17) can
therefore be approximated by
\begin{equation}
\epsilon^0_q \approx \frac{0.275sz}{0.24-0.16\ln q},\quad q\rightarrow0
\end{equation}
for a square lattice. This spectrum is very "hard" when comparing
with the magnon's soft mode $\epsilon_q\propto q$ at small $q$. For example,
we consider a system with lattice size of $N = 10^{10}$,
the smallest value for $q$ is about $q\approx 10^{-10}\pi$ and we have
energy $\epsilon^0_q \approx 0.07sz$. Comparing this value with the 
corresponding magnon energy $\epsilon_q\approx 10^{-10}sz$, we conclude
that the energy spectrum of Eq.~(17) is "nearly gapped" in a square lattice.
We also notice that the largest energy in a square lattice 
$\epsilon^0_q\approx 2.79sz$ at ${\bf q}=(\pi,\;0)$, not at 
${\bf q}=(\pi/2,\;\pi/2)$ as the case in a cubic lattice.
At ${\bf q}=(\pi/2,\;\pi/2)$, we obtain $\epsilon^0_q\approx 2.62sz$ for
the square lattice.

In summary, we have extended recently proposed variational
CCM to describe excited states and have applied to the well-studied 
Heisenberg antiferromagnets. We have obtained both the well-known 
quasiparticle excitations (magnons) by using operators $s^\pm$
and the new quasiparticle-density-wave excitations (magnon-density waves)
by using operators $s^z$. The special features and numerical values 
discussed above for the cubic and square lattices clearly indicate that 
this magnon-density-wave excitation can not be understood as a simple
sum of a pair of spin $\pm1 $ magnon excitations; they are different to
the the multi-magnon excitations as discussed in Refs. 17 and 18.
In fact, magnon-density-wave excitations share some characteristics 
with charge-density-wave excitations (plasmons) in quantum electron 
gases, with a large energy gap in three dimension and the gapless 
spectrum in two dimension; it is also interesting to know that the 
plasmon spectrum in two-dimensional system is also much "harder" than the
corresponding quasiparticle excitations \cite{dp}.
We also notice that recently modified spin-wave theories (SWT) were applied 
to finite systems with results in reasonable agreements with exact finite-size 
calculations \cite{aa,ht,mt}. As pointed out in Ref.~13, 
however, a major deficiency in this modified  SWT is the missing 
spin-zero excitations as the low-lying excitations for a finite lattice 
Heisenberg model is always triplet with spin equal to $0,\pm1$. We believe
our magnon-density-wave excitation as discussed here corresponds to
the missing branch; the energy gap in the cubic lattice and the
nearly gapped spectrum in the square lattice of Eq.~(17) reflect the
nature of long-ranged N\'eel order in the ground states of infinite
systems. In a recent experiment on the antiferromagnet MnF${}_2$
longitudinal spin fluctuations were observed and explained as two magnon 
excitations \cite{smbpr}. It will be interesting to see further experiments on
these spin-zero fluctuations in both three and two dimensional systems at very
low temperature and high energy for possible observation of magnon-density-wave
excitations as described here.

The author is grateful to former members of Department of Physics,
UMIST (before merger with University of Manchester) for their 
constant supports without which this work will not be possible.

\newpage

\begin{figure}

Fig.~1 Excitation energy spectra for the values of ${\bf q}=(0.01\pi,0)$
to $(\pi, 0)$ in a square lattice. The higher branch is for the 
magnon-density-wave excitation of Eq.~(17) and the lower one is for
the magnon excitation of Eq.~(15).

\end{figure}

\end{document}